\documentclass[12pt]{article}
\usepackage{amssymb} 
\usepackage{amsmath} 
\usepackage{color,soul}
\usepackage{amsthm}
\usepackage{epsfig}
\usepackage{graphicx, psfrag}

\begin{document}
\font\frak=eufm10 scaled\magstep1
\font\fak=eufm10 scaled\magstep2
\font\fk=eufm10 scaled\magstep3
\font\black=msbm10 scaled\magstep1
\font\bigblack=msbm10 scaled\magstep 2
\font\bbigblack=msbm10 scaled\magstep3
\font\scriptfrak=eufm10
\font\tenfrak=eufm10
\font\tenblack=msbm10


\def\biggoth #1{\hbox{{\fak #1}}}
\def\bbiggoth #1{\hbox{{\fk #1}}}
\def\sp #1{{{\cal #1}}}
\def\goth #1{\hbox{{\frak #1}}}
\def\scriptgoth #1{\hbox{{\scriptfrak #1}}}
\def\smallgoth #1{\hbox{{\tenfrak #1}}}
\def\smallfield #1{\hbox{{\tenblack #1}}}
\def\field #1{\hbox{{\black #1}}}
\def\bigfield #1{\hbox{{\bigblack #1}}}
\def\bbigfield #1{\hbox{{\bbigblack #1}}}
\def\Bbb #1{\hbox{{\black #1}}}
\def\v #1{\vert #1\vert}             
\def\ord#1{\vert #1\vert} 
\def\m #1 #2{(-1)^{{\v #1} {\v #2}}} 
\def\lie #1{{\sp L_{\!#1}}}               
\def\pd#1#2{\frac{\partial#1}{\partial#2}}
\def\pois#1#2{\{#1,#2\}}
\def\set#1{\{\,#1\,\}}             
\def\<#1>{\langle#1\rangle}        
\def\>#1{{\bf #1}}                
\def\f(#1,#2){\frac{#1}{#2}}
\def\cociente #1#2{\frac{#1}{#2}}
\def\braket#1#2{\langle#1\mathbin\vert#2\rangle} 
\def\brakt#1#2{\langle#1\mathbin,#2\rangle}           
\def\dd#1{\frac{\partial}{\partial#1}} 
\def\bra #1{{\langle #1 |}}
\def\ket #1{{| #1 \rangle }}
\def\ddt#1{\frac{d #1}{dt}}
\def\dt2#1{\frac{d^2 #1}{dt^2}}
\def\matriz#1#2{\left( \begin{array}{#1} #2 \end{array}\right) }
\def\Eq#1{{\begin{equation} #1 \end{equation}}}

\def\bw{{\bigwedge}}      
\def\hut{{\scriptstyle \land}}            
\def\dg{{\goth g^*}}                                                                                                            
\def\Cdg{{C^\infty (\goth g^*)}}
\def\poi{\{\:,\}}                           
\def\qw{\hat\omega}                
\def\FL{{\sp F}L}                 
\def\hFL{\widehat{{\sp F}L}}      
\def\XHMw{\goth X_H(M,\omega)} 
\def\XLHMw{\goth X_{LH}(M,\omega)}                  
\def\ea{\varepsilon_a}
\def\ep{\varepsilon}
\def\mitad{\frac{1}{2}}
\def\x{\times}  
\def\cinf{C^\infty} 
\def\forms{\bigwedge}                 
\def\onda{\tilde}
\def\orb{{\sp O}}

\def\a{\alpha}
\def\d{\delta}
\def\g{{\gamma }}                  
\def\G{{\Gamma}}	
\def\La{\Lambda}                   
\def\la{\lambda}                   
\def\w{\omega}                     
\def\W{{\Omega}}                   
\def\ltimes{\bowtie}

\def\roc{{\tilde{\cal R}}}                       
\def\cl{{\cal L}}                               
\def\V{{\sp V}}                                 
\def\F{{\sp F}}
\def\cv{{{\goth X}}}                    
\def\LG{\goth g}
\def\LH{\goth h}
\def\X{{{\goth X}}}                     
\def\R{{\hbox{{\field R}}}}             
\def\big R{{\hbox{{\bigfield R}}}}
\def\bbig R{{\hbox{{\bbigfield R}}}}
\def\C{{\hbox{{\field C}}}}         
\def\Z{{\hbox{{\field Z}}}}             
\def\N{{\hbox{{\field N}}}}         

\def\ima{\hbox{{\rm Im}}}                               
\def\dim{\hbox{{\rm dim}}}        
\def\End{\hbox{{\rm End}}} 
\def\Tr{\hbox{{\rm Tr}}} 
\def\tr{{\hbox{\rm\small{Tr}}}}                
\def\lin{{\hbox{Lin}}}
\def\vol{{\hbox{vol}}}  
\def\Hom{{\hbox{Hom}}}
\def\rank{{\hbox{rank}}}
\def\Ad{{\hbox{Ad}}}
\def\ad{{\hbox{ad}}}
\def\CoAd{{\hbox{CoAd}}}
\def\coad{{\hbox{coad}}}                           
\def\Rea{\hbox{Re}}                     
\def\id{{\hbox{id}}}                    
\def\Id{{\hbox{Id}}}
\def\Int{{\hbox{Int}}}
\def\Ext{{\hbox{Ext}}}
\def\Aut{{\hbox{Aut}}}
\def\Card{{\hbox{Card}}}
\def\SODE{{\small{SODE }}}
\newcommand{\bea}{\begin{eqnarray}}
\newcommand{\eea}{\end{eqnarray}}

\def\R{\mathbb{R}}
\def\ba{\begin{eqnarray}}
\def\ea{\end{eqnarray}}
\def\be{\begin{equation}}
\def\ee{\end{equation}}
\def\Eq#1{{\begin{equation} #1 \end{equation}}}
\def\R{\Bbb R}
\def\C{\Bbb C}
\def\Z{\Bbb Z}
\def\a{\alpha}                  
\def\b{\beta}                   
\def\g{\gamma}                  
\def\d{\delta}                  
\def\bra#1{\langle#1|}
\def\ket#1{|#1\rangle}
\def\goth #1{\hbox{{\frak #1}}}
\def\<#1>{\langle#1\rangle}
\def\cotg{\mathop{\rm cotg}\nolimits}
\def\Map{\mathop{\rm Map}\nolimits}
\def\wt{\widetilde}
\def\const{\hbox{const}}
\def\grad{\mathop{\rm grad}\nolimits}
\def\Div{\mathop{\rm div}\nolimits}
\def\braket#1#2{\langle#1|#2\rangle}
\def\Erf{\mathop{\rm Erf}\nolimits}
\def\matriz#1#2{\left( \begin{array}{#1} #2 \end{array}\right) }
\def\Eq#1{{\begin{equation} #1 \end{equation}}}
\def\deter#1#2{\left| \begin{array}{#1} #2 \end{array}\right| }
\def\pd#1#2{\frac{\partial#1}{\partial#2}}
\def\til{\tilde}

\def\la#1{\lambda_{#1}}
\def\teet#1#2{\theta [\eta _{#1}] (#2)}
\def\tede#1{\theta [\delta](#1)}
\def\N{{\frak N}}
\def\GR{{\cal G}}
\def\Wei{\wp}

\def\frac#1#2{{#1\over#2}} \def\pd#1#2{\frac{\partial#1}{\partial#2}}
\def\matrdos#1#2#3#4{\left(\begin{matrix}#1 & #2 \cr          
                                 #3 & #4 \cr\end{matrix}\right)}

\newtheorem{teor}{Teorema}[section]
\newtheorem{cor}{Corolario}[section]
\newtheorem{prop}{Proposici\'on}[section]
\newtheorem{note}[prop]{Note}
\newtheorem{definicion}{Definici\'on}[section]
\newtheorem{lema}{Lema}[section]
\theoremstyle{plain}
\newtheorem{theorem}{Theorem}
\newtheorem{corollary}{Corollary}
\newtheorem{proposition}{Proposition}
\newtheorem{definition}{Definition}
\newtheorem{lemma}{Lemma}

\def\Eq#1{{\begin{equation} #1 \end{equation}}}
\def\R{\Bbb R}
\def\C{\Bbb C}
\def\Z{\Bbb Z}
\def\mp#1{\marginpar{#1}}

\def\la#1{\lambda_{#1}}
\def\teet#1#2{\theta [\eta _{#1}] (#2)}
\def\tede#1{\theta [\delta](#1)}
\def\N{{\frak N}}
\def\Wei{\wp}
\def\Hil{{\cal H}}

\font\frak=eufm10 scaled\magstep1

\def\bra#1{\langle#1|}
\def\ket#1{|#1\rangle}
\def\goth #1{\hbox{{\frak #1}}}
\def\<#1>{\langle#1\rangle}
\def\cotg{\mathop{\rm cotg}\nolimits}
\def\cotanh{\mathop{\rm cotanh}\nolimits}
\def\arctanh{\mathop{\rm arctanh}\nolimits}
\def\wt{\widetilde}
\def\const{\hbox{const}}
\def\grad{\mathop{\rm grad}\nolimits}
\def\Div{\mathop{\rm div}\nolimits}
\def\braket#1#2{\langle#1|#2\rangle}
\def\Erf{\mathop{\rm Erf}\nolimits}

\centerline{\Large \bf A nonlinear superposition rule for solutions}
\medskip
 \centerline{\Large \bf of the
  Milne--Pinney equation}
\vskip 0.75cm

\centerline{ Jos\'e F. Cari\~nena
and Javier de Lucas}
\vskip 0.5cm

\centerline{Departamento de  F\'{\i}sica Te\'orica, Facultad de Ciencias,}
\medskip
\centerline{Universidad de Zaragoza,}
\medskip
\centerline{50009 Zaragoza, Spain.}
\medskip

\vskip 1cm

\begin{abstract}  
A superposition rule for two solutions of a Milne--Pinney equation is derived. 
\end{abstract}

\section{Introduction}

The search for explicit solutions of systems of first-order differential equations is not
an easy task and there exist only  few cases in which we can express in a compact
way the form of the general solution. The determination of particular solutions
can be very useful in the reduction of the original problem to a simpler one.
  In the simplest cases of linear systems
the {\sl linear superposition principle\/} plays a relevant r\^ole and we are able to 
express the general solution as an arbitrary linear combination with real
coefficients of a fundamental set of solutions. There are also cases of
nonlinear  (systems of) differential equations for which a nonlinear
superposition rule does
exist, the best known one being the Riccati equation
\begin{equation}\label{ricceq}
\frac{dx}{dt}=b_0(t)+b_1(t)x+b_2(t)x^2\,,
\end{equation}
which  is very often used in many  fields of mathematics, control theory and
theoretical physics (see for instance \cite{PW,{CMN}}  and
references therein) and its importance in this field has been
increasing since Witten's introduction of supersymmetric Quantum
Mechanics \cite{Witten81}.

 In fact, if three particular solutions of (\ref{ricceq}) 
are known,  $x_1,x_2$ and  $x_3$, the general solution can be found from 
the cross ratio relation
$$
\frac{x-x_1}{x-x_2}:\frac{x_3-x_1}{x_3-x_2}=k\ ,\qquad k\in\mathbb{R},
$$
which provides us a nonlinear superposition rule:
$$
x=\phi(x_1,x_2,x_3,k)=\frac {k\, x_1(x_3-x_2)+x_2(x_1-x_3)}{k\,(x_3-x_2)+(x_1-x_3)}\ .
$$

The general theory for nonlinear superposition rules was developed by Lie
\cite{LS} and it has been more recently revisited from a geometric perspective 
\cite{CGM00,{CGM07}}.

Another nonlinear equation with relevance in physics is the today called 
Milne--Pinney equation:
\begin{equation}
\ddot x=-\omega^2(t)x+\frac k{x^3}\,, \qquad k>0\,.\label{MPeq}
\end{equation}
This equation is defined on $\mathbb{R}-\{0\}$ and it is invariant under
parity, i.e. if $x(t)$ is a solution then $\bar x(t)=-x(t)$ is a solution too.
That means that it is enough to restrict ourselves to the $x>0$ case.

Equation (\ref{MPeq}) models many physical problems such as
propagation of laser beams in nonlinear media, plasma dynamics, or the mean
field dynamics for Bose--Einstein condensates through the so-called  
Gross--Pitaevskii equation.
 It was introduced by Ermakov as a way for finding a first-integral of the 
corresponding time-dependent
 harmonic oscillator described by 
\begin{equation}
\ddot y+\omega^2(t)y=0 \label{TDFHOeq}
\end{equation}  by means of the Ermakov 
system \cite{Er80}  made up by  the two equations (\ref{MPeq}) and
(\ref{TDFHOeq}). 
This Ermakov system admits 
the well-known Ermakov invariant
\begin{equation}
I(x,y,v_x,v_y)= k\left(\frac{y}{x}\right)^2+(xv_y-yv_x)^2\,,\label{Erminv}
\end{equation}
from  where one can see that the general solution $x(t)$ of Milne--Pinney equation
can be expressed in terms of two solutions $y_1$ and $y_2$  of the associated 
harmonic oscillator 
as follows \cite{P50}:
\begin{equation}\label{OldSR}
x=\frac{\sqrt{2}}{|W|}\sqrt{C_1y_1^2+C_2y_2^2\pm\sqrt{4C_1C_2-kW^2}y_1y_2}\,,
\end{equation}
where $W$ is the Wronskian of the two solutions of (\ref{TDFHOeq}), $W=y_1\dot y_2-y_2\dot y_1$,
which is constant, 
and  $C_1$ and $C_2$ are non-negative constants such that $4C_1C_2\geq kW^2$. In this way, as
$$
(\sqrt{C_1}y_1-\sqrt{C_2}y_2)^2\geq 0
$$
then $C_1y_1^2+C_2y_2^2\geq \sqrt{4 C_1C_2}y_1 y_2>\sqrt{4 C_1C_2-kW^2}y_1 y_2$ and thus
the solutions given by (\ref{OldSR}) are real.

In a similar way, it has recently been shown in \cite{CLR08} that there exists also a
superposition rule involving three solutions of a related Riccati equation 
for giving the general solution of (\ref{MPeq}). 

 A generalisation of this Ermakov system was proposed in \cite{RR79a,{RR80},{SaCa82},{WS81}}:
\begin{equation}
\left\{\begin{array}{rcl}
\ddot{x}&=&{\displaystyle\frac{1}{x^3}}f(y/x)-\omega^2(t)x\cr&&\cr
\ddot{y}&=&{\displaystyle\frac{1}{y^3}}g(y/x)-\omega^2(t)y
\end{array}\right.,
\end{equation}
and a new
invariant was obtained for this generalised Ermakov system:
\begin{equation}
  I(x,y,v_x,v_y)= (xv_y-yv_x)  ^2+2\,\int^{x/y}\left[-\frac 1{u^3}\, f\left(\frac 1u\right)+
  u\,g\left(\frac 1u
\right)\right]\,du\,.\label{genErminv}
\end{equation}
This  generalised Ermakov system reduces to the standard one for   $f(u)=k$
and $g(u)=0$. In this case (\ref{genErminv}) becomes (\ref{Erminv}).
\section{A new superposition rule for the Milne--Pinney equation.}
Our aim in this section is to show that there exists 
 a superposition rule for the Milne--Pinney equation (\ref{MPeq}) \cite{P50,Mil30, CLR07a}
in terms of a pair of its particular solutions.

In fact, one see from (\ref{genErminv}) that in the
particular case of
$f=g=k$, if  a particular solution $x_1$ is known,  there is a $t$-dependent
constant
 of motion  for the Milne--Pinney
equation
given by (see e.g.  \cite{CLR07a}):
\begin{equation}\label{C1}
I_1=(x_1\dot x-\dot x_1x)^2+k\left[\left(\frac{x}{x_1}\right)^2+\left(\frac{x_1}{x}\right)^2\right]\,.
\end{equation}

If another particular solution $x_2$ of (\ref{MPeq})  is given, then
we have another $t$-dependent constant of motion 
\begin{equation}\label{C2}
I_2=(x_2\dot x-\dot x_2x)^2+k\left[\left(\frac{x}{x_2}\right)^2+\left(\frac{x_2}{x}\right)^2\right]\,.
\end{equation}
Moreover, the two solutions $x_1$ and   $x_2$ provide us a function of $t$ which
is constant and generalises the Wronskian $W$ of two solutions of (\ref{MPeq}):
\begin{equation}\label{C3}
I_3=(x_1\dot x_2-x_2\dot x_1)^2+k\left[\left(\frac{x_2}{x_1}\right)^2+\left(\frac{x_1}{x_2}\right)^2\right]\,.
\end{equation}

Remark that for any real number $\alpha$ the inequality
$(\alpha-1/\alpha)^2\geq 0$ implies 
$$\alpha^2+\frac 1{\alpha^2}\geq 2\,,$$
and the equality sign is valid if and only if $|\alpha|=1$, 
$$\alpha^2+\frac
1{\alpha^2}= 2\Longleftrightarrow \ |\alpha|=1\,.
$$
Therefore, as we have considered $k>0$, we see that $I_i\geq 2\,k$, 
for $i=1,2,3$.  
Moreover, as the solutions  $x_1(t)$ and $x_2(t)$ are different
 solutions of the Milne--Pinney equation, it turns out that $I_3>2k$.

The knowledge of the two constants of motion  $I_1$ and $I_2$, together with
 the constant value of
$I_3$ 
for a pair of solutions of (\ref{MPeq}),
can be used to obtain the superposition rule for such a differential
equation. In fact, 
given two solutions $x_1$ and $x_2$  of (\ref{MPeq}), first  the constant of motion 
 (\ref{C2}) allows us to write an explicit expression for $\dot x$ in terms of  $x, x_2$ and $I_2$:
\[
\dot x=\dot x_2 \frac{x}{x_2}\pm\sqrt{-k\frac{x^2}{x_2^4}+I_2\frac{1}{x_2^2}-k\frac{1}{x^2}}\,,
\]
and  using  such  expression in (\ref{C1}) we see after a careful computation  
 that $x^2$ satisfies the following fourth degree equation
\begin{multline}\label{bicua}
(I_2^2-4k^2)x_1^4-2(I_1I_2-2I_3 k)x_1^2x_2^2+(I_1^2-4 k^2)x_2^4
-\\-2((I_2I_3-2 I_1 k)x_1^2+(I_1 I_3-2 I_2k)x_2^2) x^2+(I_3^2-4 k^2) x^4=0\,,
\end{multline}
where use has been made of the constant value of  $I_3$ for two solutions of
the Milne--Pinney equation.

Therefore,  we can obtain from (\ref{bicua}) the expression for the square of
the  solutions of (\ref{MPeq}) in terms of any pair of its particular positive solutions by means of a superposition rule 
\begin{equation}
x^2=\lambda_1x_1^2+\lambda_2x_2^2\pm
2\sqrt{\lambda_{12}[-k (x_1^4+x_2^4)+I_3\,x_1^2x_2^2\,]},\label{SRxdos}
\end{equation}
where the constants $\lambda_1$ and $\lambda_2$ are given by
\begin{eqnarray}\label{lambdas}
\lambda_1=\frac{I_2 I_3-2 I_1 k}{I_3^2-4 k^2},\qquad \lambda_2=\frac{I_1 I_3-2 I_2 k}{I_3^2-4 k^2}\,,
\end{eqnarray}
and $\lambda_{12}$ is a constant which depends on $\lambda_1$, $\lambda_2$, $k$
and the constant value of $I_3$ as follows:
\begin{equation}
\lambda_{12}=\lambda_{12}( \lambda_1, \lambda_2; I_3, k)=
\frac{\lambda_1\lambda_2 I_3+k(-1+\lambda_1^2+\lambda_2^2)}{I_3^2-4
  k^2}=\varphi( I_1, I_2; I_3, k)\,,
\end{equation}
where the function $\varphi$ is given by
\begin{equation}
\varphi(I_1, I_2; I_3, k)=\frac{I_1 I_2 I_3-(I_1^2+I_2^2+I_3^2)k+4k^3}{(I_3^2-4 k^2)^2}\,.
\end{equation}

It is important to remark that  if $\lambda_1<0$ then $\lambda_2>0$ and if
$\lambda_2<0$ then $\lambda_1>0$, i.e. if $\lambda_1<0$ then $I_2I_3<2I_1k$, and  thus $I_2<2kI_1/I_3$. Therefore,
$\lambda_2(I_3^2-4k^2)=I_1I_3-2k I_2>I_1I_3-4k^2 I_1/I_3=I_1(I_3^2-4k^2)>0$, and thus, as $I_3>2k$, $\lambda_2>0$. Similarly we obtain that $\lambda_2<0$ implies $\lambda_1>0$. 

The parity invariance of (\ref{MPeq}) is displayed by (\ref{SRxdos}) which
gives us the solutions 
\begin{equation}\label{SR}
x^2=\lambda_1x_1^2+\lambda_2x_2^2\pm
2\sqrt{\lambda_{12}[-k (x_1^4+x_2^4)+I_3\,x_1^2x_2^2\,]}\,.
\end{equation}

In order for $x^2$ to be a positive  real number giving rise to a real solution
 of the Milne--Pinney equation the constants $\lambda_1$ and $\lambda_2$ in
 the 
preceding  expression should satisfy some additional restrictions. 
In particular, they must be such that $$\lambda_{12}[-k
(x_1^4(0)+x_2^4(0))+I_3\,x_1^2(0)x_2^2(0)\,]\geq 0$$ and $$\lambda_1x_1^2(0)+\lambda_2x_2^2(0)\pm
2\sqrt{\lambda_{12}[-k (x_1^4(0)+x_2^4(0))+I_3\,x_1^2(0)x_2^2(0)\,]}>0.$$ If
these conditions are satisfied, then  differentiating (\ref{SR}) 
in $t=0$ for $x_1=x_1(t)$ and $x_2=x_2(t)$ solutions of (\ref{MPeq}), it can be
checked out that $\dot x(0)$  is also a real constant. As $x(t)$ is a solution
with real initial conditions then $x(t)$ given by (\ref{SR}) is real in an interval of $t$ and thus all the obtained conditions will be valid in an interval of $t$. 

If we take into account that we have consider $x_2> 0$, we can simplify  the study of such restrictions by  writing (\ref{SR}) in terms of the variables $x_2$ and $z=(x_1/x_2)^2$ as
\begin{equation}
x^2=x_2^2\left(\lambda_1z+\lambda_2\pm
2\sqrt{\lambda_{12}[-k (z^2+1)+I_3\,z\,]}\right)\,,
\end{equation}
and  the preceding  conditions turn out to be  $\lambda_{12}[-k (z^2+1)+I_3\,z\,]\geq 0$ and $\lambda_1z+\lambda_2\pm
2\sqrt{\lambda_{12}[-k (z^2+1)+I_3\,z\,]}>0$. 

Next, in order to get  $\lambda_{12}[-k (z^2+1)+I_3\,z\,]\geq0$, we 
first note that 
 this expression  is not definite because its discriminant is 
$\lambda_{12}^2(I_3^2-4 k^2)\geq 0$, and this restricts the possible values of
$\lambda_1$ and 
$\lambda_2$ for a given $z$.
With this aim we define the polynomial $P(z)$ given by
\begin{equation}
P(z)=-k(z^2+1)+I_3\,z,
\end{equation}
with roots 
\begin{equation}
z=z_\pm=\frac{I_3\pm\sqrt{I_3^2-4k^2}}{2k},
\end{equation}
which  can be written in terms of the variable $\alpha_3=I_3/2k$ as
\begin{equation}
z_\pm=\alpha_3\pm\sqrt{\alpha_3^2-1}.
\end{equation}

As $\alpha_3>1$, then $\alpha_3>\sqrt{\alpha_3^2-1}>0$ and thus
$z_\pm>0$. The sign of the polynomial $P(z)$ is  displayed  in Fig. 1.

\begin{figure}[ht!]
\centerline{
\psfrag{a}{$t\equiv x_1=\sqrt{\alpha_3}x_2$}
\psfrag{b}{$r\equiv x_1=\sqrt{z_-}x_2$}
\psfrag{c}{$s\equiv x_1=\sqrt{z_+}x_2$}
\psfrag{e}{$x_1$}
\psfrag{f}{$x_2$}
\psfrag{g}{}
\psfrag{i}{$A(-)$}
\psfrag{j}{$B(+)$}
\psfrag{k}{$B(+)$}
\psfrag{l}{$A(-)$}
\psfrag{m}{}
\includegraphics[width=8cm]{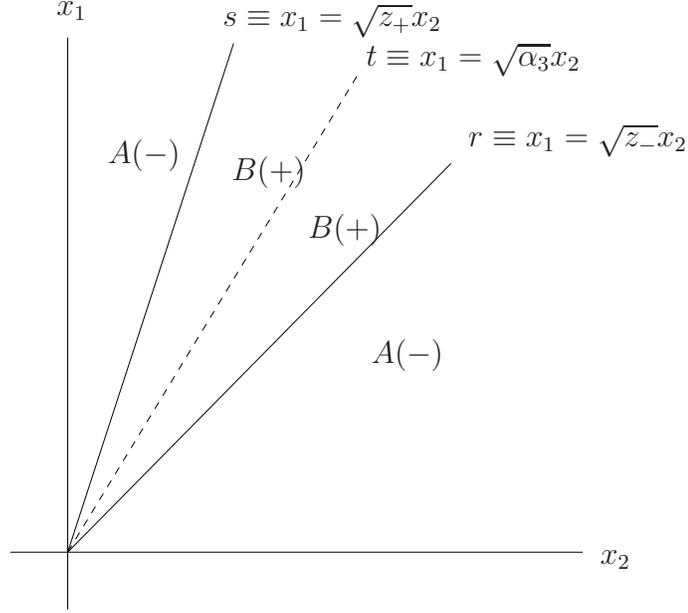}}
\caption{Sign of the polynomial $P(x_1,x_2)$.}
\end{figure}

The region  $\mathbb{R}_+\times \mathbb{R}_+$ splits into three regions, 
$$A=\{(x_1,x_2)\in \mathbb{R}_+\times \mathbb{R}_+\mid x_1>\sqrt{z_+}\, x_2\}\bigcup 
\{(x_1,x_2)\in \mathbb{R}_+\times \mathbb{R}_+\mid x_1<\sqrt{z_-}\, x_2\}\,,$$
$$B=\{(x_1,x_2)\in \mathbb{R}_+\times \mathbb{R}_+\mid \sqrt{z_-}x_2<
x_1<\sqrt{z_+} x_2\}$$  separated by the region 
 $$C=\{(x_1,x_2)\in \mathbb{R}_+\times \mathbb{R}_+\mid x_1=\sqrt{z_+}\, x_2\}\bigcup 
\{(x_1,x_2)\in \mathbb{R}_+\times \mathbb{R}_+\mid x_1=\sqrt{z_-}\, x_2\}$$
of 
 the straight lines $x_1=\sqrt{z_+}x_2$ and  $x_1=\sqrt{z_-}x_2$.  The condition to make $\lambda_{12}P(z)$ non-negative
in region $A$ where $P$ takes negative values is 
 to choose $\lambda_1$ and
 $\lambda_2$ such that $\lambda_{12}(\lambda_1,
 \lambda_2, I_3, k)\leq 0$. Similarly, as $P$ is positive in region $B$
 we have to choose $\lambda_1$ and $\lambda_2$ such that
 $\lambda_{12}(\lambda_1, \lambda_2, I_3, k)\geq 0$. Finally as $P$ vanishes in 
region $C$ there is no restriction on the coefficients
 $\lambda_1$ and $\lambda_2$.

Once we have stated the conditions for $\lambda_{12}P(z)$ to be
non-negative we still have to impose the condition 
\begin{equation}\label{Con4}
\lambda_1z+\lambda_2\pm
2\sqrt{\lambda_{12}[-k (z^2+1)+I_3\,z\,]}> 0.
\end{equation}

In order to study these conditions we study the sign of the polynomial
\begin{equation}\label{SecondPol}
\begin{aligned}
P_{I_3,k}(z,\lambda_1,\lambda_2)&=(\lambda_1z+\lambda_2)^2-4\lambda_{12}[-k(z^2+1)+I_3z]\\
&=\frac{4P(z)I_3}{I_3^2-4 k^2}+\left(a\lambda_1+b\lambda_2\right)^2\,,
\end{aligned}
\end{equation}
where
\begin{equation}
a=\sqrt{-\frac{4P(z)k}{I_3^2-4k^2}+z^2},\qquad b=\sqrt{1-\frac{4P(z)k}{I_3^2-4k^2}}.
\end{equation}

As we remarked before,  $\lambda_1,\lambda_2$  cannot be both negative. Let $K$
denote the set 
$K=\mathbb{R}^2-\{(\lambda_1,\lambda_2)\in\mathbb{R}^2\mid \lambda_1<0, \lambda_2<0 \}$ and 
consider three cases:
\begin{enumerate}
\item If $(x_1,x_2)\in A$, then as $P(z)\leq 0$,  it must be $\lambda_{12}\leq
  0$ in order to
 satisfy $\lambda_{12}P(z)\geq 0$. In this case, if $K_1$ and $K_2$ are
 the sets 
\begin{equation}
\begin{aligned}
K_1=\left\{(\lambda_1,\lambda_2) \in K ;  \sqrt{-\frac{4P(z)I_3}{I_3^2-4k^2}}>|a\lambda_1+b\lambda_2|\right\}\,,\\
K_2=\left\{(\lambda_1,\lambda_2) \in K ; \sqrt{-\frac{4P(z)I_3}{I_3^2-4k^2}}<|a\lambda_1+b\lambda_2|\right\}\,.
\end{aligned}
\end{equation}
 We find   the following particular
subcases
\begin{enumerate}
\item  If   $(\lambda_1,\lambda_2)\in K_1$, then $P_{I_3,k}(z,\lambda_1,\lambda_2)>0$.
\item  If   $(\lambda_1,\lambda_2)\in K_2$ then $P_{I_3,k}(z,\lambda_1,\lambda_2)<0$,
\end{enumerate}
that can be summarised  by means of Figure 2.

\begin{figure}[ht!]
\centerline{\psfrag{a}{$\lambda_1$}
\psfrag{b}{$\lambda_2$}
\psfrag{c}{$K_1$}
\psfrag{d}{$K_2$}
\includegraphics[width=8cm]{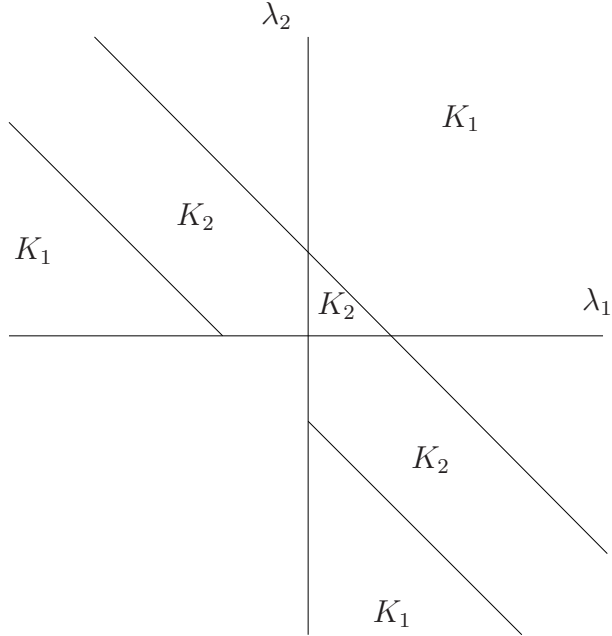}}
\caption{Sign of the polynomial $P_{I_3,k}(z,\lambda_1,\lambda_2)$ in $K$.}
\end{figure}
\item If $(x_1, x_2)\in B$, as $P(z)$ is positive, then  $\lambda_{12}      $ must
  also be positive, $\lambda_{12}\geq 0$. Thus for $(\lambda_1, \lambda_2)\in K_1\cup K_2$, $P_{I_3,k}(z,\lambda_1,\lambda_2)>0$. 
\item If $(x_1,x_2)\in C$, then for $(\lambda_1, \lambda_2)\in K_1\cup K_2$, $P_{I_3,k}(z,\lambda_1,\lambda_2)>0$.
\end{enumerate}

In those cases in which $P_{I_3,k}(z,\lambda_1,\lambda_2)>0$ we can assert that $|\lambda_1z+
\lambda_2 |>2\sqrt{\lambda_{12}[-k (z^2+1)+I_3\,z\,]}$,
but we still have to impose  that $\lambda_1z+\lambda_2>0$ for (\ref{Con4}) to
be
 positive. Nevertheless, this is very simple, because 
if such a pair $(\lambda_1 ,\lambda_2)$ does not  satisfy 
 $\lambda_1z+\lambda_2>0$, the pair of opposite elements
 $(-\lambda_1,-\lambda_2)$ actually does it,  the other conditions being
 invariant 
under the change $\lambda_i\rightarrow -\lambda_i$ with $i=1,2$.

In those cases in which $P_{I_3,k}(z,\lambda_1,\lambda_2)<0$ we can assert that $|\lambda_1z+
\lambda_2|<2\sqrt{\lambda_{12}[-k (x_1^4+x_2^4)+I_3\,x_1^2x_2^2\,]}$, and in
this case the unique valid superposition rule is
\begin{equation}
x=|x_2|\left(\lambda_1z+\lambda_2+
2\sqrt{\lambda_{12}[-k (z^2+1)+I_3\,z\,]}\right)^{1/2}\,,
\end{equation}
which is equivalent to
\begin{equation}
x=\left(\lambda_1x_1^2+\lambda_2x_2^2+
2\sqrt{\lambda_{12}[-k (x_1^4+x_2^4)+I_3\,x_2^2x_1^2\,]}\right)^{1/2}\,.
\end{equation}

It is important to notice that the given conditions for $\lambda_1, \lambda_2$
are used to obtain real solutions of the Milne--Pinney equation and if 
 no restriction were
   considered
 we would had obtained real and imaginary solutions for the real Milne--Pinney equation.

Expression (\ref{SR}) provides us  a (as far as we know, new)
 superposition rule for the positive solutions of the Pinney equation (\ref{MPeq})
in terms of two of its independent particular positive solutions. 
Therefore, once two particular solutions of the equation (\ref{MPeq}) are known, 
we can write its general solution.
Note also that, because
of the parity symmetry of 
 (\ref{MPeq}), the superposition (\ref{SR}) can be used with positive and
 negative solutions. In all these ways we obtain if $k> 0$ non-vanishing  solutions of
 (\ref{MPeq}). It is also  to 
be remarked that the same  procedure as before
 can be repeated in the case $k<0$ to obtain similar results.

A similar superposition rule works for negative solutions of Milne--Pinney
equation (\ref{MPeq}): 
\begin{equation}\label{SR2n}
x=-\left(\lambda_1x_1^2+\lambda_2x_2^2\pm
2\sqrt{\lambda_{12}(-k (x_1^4+x_2^4)+I_3\,x_1^2x_2^2)}\right)^{1/2}.
\end{equation}
where once  again 
 $x_1$ and $x_2$ are arbitrary  solutions.
\section{Relations between the new and the known superposition rule.}
We can now compare both superpositions rules (\ref{OldSR}) and (\ref{SR}) and
check that actually the latter reduces to the former one  when $x_1$ and $x_2$ are
obtained from solutions $y_1$ and $y_2$ of the associated harmonic oscillator
equation.

Let $y_1$ and $y_2$ be two solutions of (\ref{TDFHOeq}) and $W$ its Wronskian.  Consider 
the 
 two particular positive solutions of the Milne--Pinney-equation $x_1(t)$ and
 $x_2(t)$
given by 
\begin{equation}\label{Change}
\begin{aligned}
x_1&=\frac{\sqrt{2}}{|\, W\,|}\sqrt{C_1y_1^2+C_2y_2^2},\\
x_2&=\frac{\sqrt{2}}{|\, W\,|}\sqrt{C_2y_1^2+C_1y_2^2},
\end{aligned}
\end{equation}
where $C_1< C_2$ and we additionally impose 
\begin{equation}
 4C_1C_2=kW^2\,.\label{addcond}
\end{equation}  

      $I_3$ given by (\ref{C3}) for the two particular solutions of the
Milne--Pinney equation can then be expressed  as a function of the solutions 
$y_1$ and $y_2$ of   
 the time-dependent harmonic oscillator and its Wronskian $W$. After a long computation $I_3$ 
turns out to be
\begin{equation}\label{I3}
I_3=\frac{4(C_1^2+C_2^2)}{W^2}\,,
\end{equation}
and then  using the explicit form (\ref{Change}) of the particular solutions
and
taking into account (\ref{I3}) in (\ref{SR}) we obtain that 

\begin{multline}
\lambda_1 x_1^2+\lambda_2x_2^2\pm 2\sqrt{\lambda_{12}(-k(x_1^4+x_2^4)+I_3 x_1^2x_2^2)}=
{\displaystyle\frac{2}{W^2}}(C_1\lambda_1+C_2\lambda_2)y_1^2\\ +(C_1\lambda_2+C_2\lambda_1)y_2^2)
\pm{\displaystyle\frac{2}{W^2}}\sqrt{4(C_1\lambda_1+C_2\lambda_2)(C_1\lambda_2+C_2\lambda_1)-kW^2}y_1y_2.
\end{multline}
Consequently,  from the superposition rule (\ref{SR}) we recover  expression 
(\ref{OldSR}):
\begin{equation}\label{SR2}
x=\frac{\sqrt{2}}{|\, W\,|}\sqrt{\mu_1y_1^2+\mu_2y_2^2\pm\sqrt{4\mu_1\mu_2-k W^2}y_1 y_2},
\end{equation}
where now
$$
\left\{\begin{aligned}
\mu_1&=(C_1\lambda_1+C_2\lambda_2),\\
\mu_2&=(C_1\lambda_2+C_2\lambda_1).
\end{aligned}\right.
$$
Once we have stated the superposition rule we still have to analyse the
possible values of $\lambda_1$ and $\lambda_2$ that we can use in this case. If
we use
 the expression          (\ref{I3}) we obtain after a short  calculation 
   the following 
values $z_\pm$ 
\begin{equation}
z_+=\frac{4C_2^2}{kW^2},\quad z_-=\frac{4C_1^2}{kW^2}.
\end{equation}
Now if we write    $y_1^2$ and $y_2^2$ 
 in terms of $x_1^2, x_2^2$ and $W$ from the system (\ref{Change})
we obtain 
\begin{equation}
\frac 1{C_1^2-C_2^2}
\left(\begin{array}{cc}
C_1&-C_2\\-C_2&C_1
\end{array}\right)
\left(\begin{array}{c}x_1^2\\x_2^2\end{array}\right)
=\left(\begin{array}{c}y_1^2\\y_2^2 \end{array} \right).
\end{equation}

Therefore, as $C_2>C_1$ the condition of being $y_1^2$ and 
$,y_2^2     $ being positive is
\begin{equation}\left\{
\begin{aligned}
C_1x_1^2\leq C_2x_2^2\\
C_2x_1^2\geq C_1x_2^2\\
\end{aligned}\right.
\end{equation}
and it is verified if $x_1^2/x_2^2
\leq C_2/C_1=4C_2^2/kW^2=z_+$ and $x_1^2/x_2^2     \geq
C_1/C_2=4C_1^2/kW^2=z_-$,
because of (\ref{addcond}).
 Thus, $(x_1, x_2)\in B$ and therefore      the only restrictions for
 $\lambda_1, \lambda_2$ are $\lambda_{12}\geq 0$ and
 $\lambda_1x_1^2+\lambda_2x_2^2\geq 0$. Obviously, by means of the change of
 variables (\ref{Change}) this last expression is equivalent to
 $\mu_1y_1^2+\mu_2y_2^2\geq 0$ and thus $\mu_1$ and $\mu_2$ can not be
 simultaneously negative. Furthermore,
      $\lambda_{12}(I_3^2-4k^2)=4\mu_1\mu_2-kW^2$. As we have said that
      $\lambda_{12}\geq 0$ then $4\mu_1\mu_2\geq kW^2$, i.e. $\mu_1\mu_2$ is
      positive and thus, $\mu_1$ and $\mu_2$               are positive.
In this way we recover  the usual constants of the known superposition rule of the Milne--Pinney equation in terms of solutions of an harmonic oscillator.
\section{A simple example of the new superposition rule.}
As a particularly simple but interesting application, we can  use the
superposition rule
(\ref{SR})  for the Milne--Pinney equation in order to obtain the general
solution
 of the constant unity frequency case:
\begin{eqnarray}\label{example}
\ddot{x}=\frac{k}{x^3}-x\,.
\end{eqnarray}

As indicated by  Pinney in \cite{P50}, the superposition rule (\ref{SR}) 
gives us the  general solution in terms of solutions of the corresponding 
harmonic oscillator
\begin{eqnarray}\label{harosc}
\begin{aligned}
\ddot{y}=-y.
\end{aligned}
\end{eqnarray}

Choosing two positive
 constants $C_1$ and $C_2$ such that  $4C_1C_2=kW^2$ and $C_1< C_2$,
 we obtain the following  two independent  particular solutions of the
 Pinney-equation:
\begin{equation}
\begin{aligned}
x_1&=\frac{\sqrt{2}}{|\, W\,|}\sqrt{C_1y_1^2+C_2y_2^2},\\
x_2&=\frac{\sqrt{2}}{|\, W\,|}\sqrt{C_2y_1^2+C_1y_2^2}.
\end{aligned}
\end{equation}
If we take, for instance,  $y_1=\cos t$ and $y_2=\sin t$, for which $W=1$, 
 the following  solutions for the Milne--Pinney equation
are obtained:
\[
\begin{aligned}
x_1&={\sqrt{2}}\sqrt{C_1\cos^2(t)+C_2\sin^2(t)},\\
x_2&={\sqrt{2}}\sqrt{C_2\cos^2(t)+C_1\sin^2(t)}.
\end{aligned}
\]

We can use these two expressions in (\ref{SR}) for obtaining
 the general solution  of the Milne--Pinney equation. If we choose $C_1=\sqrt{k}/4$ and $C_2=\sqrt{k}$ then we obtain the general solution for the Milne--Pinney equation with constant frequency:
\begin{multline}
x(t)=\frac{\sqrt{k}}{2}\left(5(\lambda_1+\lambda_2)+3(\lambda_1-\lambda_2)\cos(2t)\right.\\+
2\sqrt{(4\lambda_1+\lambda_2)(\lambda_1+4\lambda_2)-4}\sin(2t))^{1/2}\,.
\end{multline}
         Obviously, as we said before, $\lambda_1$ and $\lambda_2$ are such that $\lambda_{12}\geq 0$ and $\lambda_1x_1^2+\lambda_2x_2^2\geq 0$.

\section{Conclusions and Outlook.}
In summary, 
we have obtained a new superposition rule for the Milne--Pinney equation in
terms of a set of two of its   particular solutions. Also, we have related the
new superposition rule with the one proposed in \cite{P50} for solutions
of the related equation (\ref{TDFHOeq}). Finally,  a particular example in
which we use this new superposition rule to obtain the general solution for the
corresponding Milne--Pinney equation is given.

Another  interesting question to be solved later on is why the obtained
superposition rule for the Milne--Pinney equation and the already known one
just depend on the variables $x_1$ and $  x_2$ and not      on the $v_1$ and $v_2$ variables as one would expect by the theory of Lie systems. This fact is not an isolated result and can be found also in other superposition rules for second order differential equations, see \cite{CLR07a}.

\subsection*{Acknowledgements}
Partial financial support by research projects MTM2006-10531 and E24/1 (DGA)
 are acknowledged. JdL also acknowledges
 a F.P.U. grant from  Ministerio de Educaci\'on y Ciencia.

\end{document}